\documentclass[3p,times,twocolumn]{elsarticle}

\usepackage{ecrc}
\usepackage{amsmath,amssymb}

\usepackage{graphics}
\usepackage{graphicx}
\usepackage{epsfig}

\volume{00}

\firstpage{1}

\journalname{Physica B}

\runauth{S.V. Zaitsev-Zotov et al.}

\jid{physb}

\jnltitlelogo{Physica B}

\usepackage{amssymb}
\begin{document}

\begin{frontmatter}


 \title{Charge-density waves physics revealed by photoconduction}
 \author[label1,label2]{S.V. Zaitsev-Zotov\corref{cor1},}
 \author[label1]{V.F. Nasretdinova,}
  \author[label1]{V.E. Minakova}
 \address[label1]{Kotel'nikov Institute of Radio-engineering and Electronics of the RAS, 125009 Moscow, Russia} 
 \address[label2]{Moscow Institute of Physics and Technology, 141700 Dolgoprudny, Russia }
 \cortext[cor1]{Tel: +7 (495) 629 33 94; FAX:+7 (495) 629 36 78; e-mail: serzz@cplire.ru}

%\title{}


%\author{}

%\address{}

\begin{abstract}
The results of photoconduction study of the Peierls conductors are reviewed. 
%A brief review of experimental data obtained from study of the Peierls conductors by means of photoconduction is given. 
The studied materials are quasi-one-dimensional conductors with the charge-density wave: K$_{0.3}$MoO$_3$, both monoclinic and orthorhombic TaS$_3$ and also a semiconducting phase of NbS$_3$ (phase I). Experimental methods, relaxation times, effects of illumination on linear and nonlinear charge transport, the electric-field effect on photoconduction and results of the spectral studies are described. We demonstrate, in particular, that a simple model of modulated energy gap slightly smoothed by fluctuations fits the available spectral data fairly well. The level of the fluctuations is surprisingly small and does not exceed a few percent of the optical energy gap value.
\end{abstract}

\begin{keyword}
Peierls conductors \sep charge-density wave  \sep photoconduction \sep electron transport \sep collective transport \sep energy structure, van Hove singularity 
\PACS 71.45.Lr \sep  72.15.Nj  \sep 71.20.Ps \sep 72.20.Jv \sep 73.20.Mf
\end{keyword}
\end{frontmatter}

\section{Introduction}

Despite of impressive development of modern experimental methods such as ARPES, Fourier spectroscopy, femtosecond spectroscopy, scanning tunneling spectroscopy {\it etc.}, more than 150 years old photoconduction remains to be very useful method for both material characterization and study of underlying physics. Application of this method to investigation of the quasi-one dimensional (q-1D) conductors with the charge density wave (CDW) provide a unique information on various aspects of very reach physics of the Peierls conductors (see \cite{review} as a recent review).  
Here we present a brief overview of the modern state of research in this area. 

The traditional definition of photoconduction as a phenomenon in which a material becomes more conductive due to absorption of electromagnetic radiation does not work in the case of q-1D  conductors with the CDW where light absorption under certain conductions (temperature and voltage range) may make the material less conductive. So we will consider photoconduction  as a phenomenon in which a material changes its conduction in any direction due to light absorption at a {\it constant} temperature. The latter condition is essential to distinguish photoconduction and the bolometric effect where conductance changes are caused by temperature variation caused by light absorption. 

The underlying physical mechanisms of photoconduction are related to light-induced electronic transitions. Only those transitions contribute into conductance, which change the current carrier concentration or mobility. In this respect photoconduction study and dielectric spectroscopy or bolometric response study provide different information, because the latter are also sensitive to electronic transitions which do not change conductance. 

We will use here the following notation: $G$ is a sample conductance, $\delta G\equiv G(W)-G(0)$ is photoconductance, i.e. the conductance variation due to illumination, $W$ is a light intensity at the sample position, $I$, $V$, $R\equiv 1/G$ and $L$ are respectively the electric current, voltage, sample resistance and length, $E\equiv V/L$ is the electric field, applied to the sample, $2\Delta_{opt}$ and $2\Delta_{tr}$ are the optical and transport energy gaps, $D$ is the Gauss fluctuation dispersion. The spectral data discussed below are normalized to the flow of incident photons i.e. to the ratio $S(\hbar\omega)\equiv W/\hbar\omega$.

%Photoconduction is known since 1873 when it was discovered by U Smith and since there thousands of papers were devoted to this phenomana. 

\section{Experimental methods}

Photoconduction study needs a light source optically coupled to a sample. In its simplest realization, a LED capable to operate at low-temperatures can be placed in the close vicinity of the sample. 
%In this case it can be used to study temperature- and light intensity-dependent relaxation of the nonequilibrium current carriers at relatively small excitation level, in contrast to e.g. femtosecond spectroscopy which works at very high excitation levels. 
Special precautions should be undertaken to minimize the heating and capacitive coupling of the LED with the sample. Temperature dependence of the LED spectrum should also be taken into account in some special cases. As the LED-emitted light power is controlled by the current flowing through the LED, the light intensity control is also simple. This method was widely used for investigation of the temperature and time evolution of the photoconduction current in q-1D conductors \cite{zzminajetpl,zzmprl,zzminajphys,minahere}.

Spectral study is a little bit more complex. Ogawa with coauthors \cite{ogawa} used short laser pulses (200 fs) generated by OPG-OPA (optical parametric generator and optical parametric amplifier) to study the electro-optical response of K$_{0.3}$MoO$_3$. In our experiments we used the grid monochromator IKS31 with a globar or quartz lamp (the choice depends on the spectral region) as a light source, the light was modulated mechanically and the photoconductance signal was measured by a lock-in amplifier  \cite{nzzjetpl,nbs3,zzmnzreview,venerahere}. 
% (other details can be found in the original publications \cite{nzzjetpl,nbs3,zzmnzreview}). 
Note that nonlinearity of photoconduction response, $\delta G(W)$, intrinsic to all Peierls conductors studied up to now complicates usage of advantages of Fourier spectroscopy.
% to obtain photoconduction spectral data.

 %Presence of Wood's anomalies in the output spectra provided by a grid spectrometer is not completely suppressed by such a calibration due to anisotropy of the photoconduction response can be completely eliminated in the polarization measurements only.

%\section{Competitive spectral methods}
%In conclusion, it is interesting to compare photocondution with other methods which can be used for energy structure study: Fourier spectroscopy, dielectric spectroscopy, ARPES, tunneling, Raman, reflection, adsorption, bolometric response
%
%table of comparison

%\section{Notation}

\section{Temperature and electric field dependences of photoconductance}
%\section{Studied materials}
To our knowlege photoconduction was studied in four inorganic q-1D materials: the blue bronze K$_{0.3}$MoO$_3$ and transition metal trichalcogenides: monoclinic and orthorhombic TaS$_3$,  and also in NbS$_3$ (phase I). The first three materials are typical Peierls conductors with the Peierls transition temperatures $T_P=180$~K for the blue bronze, 220~K for o-TaS$_3$, and  240~K and 160~K for m-TaS$_3$.  NbS$_3$ (phase I) is considered as an anisotropic semiconductor with 1~eV optical gap value  \cite{review}. 

\subsection{Temperature dependences of photoconductance}

\begin{figure}
\includegraphics[width=7.5cm]{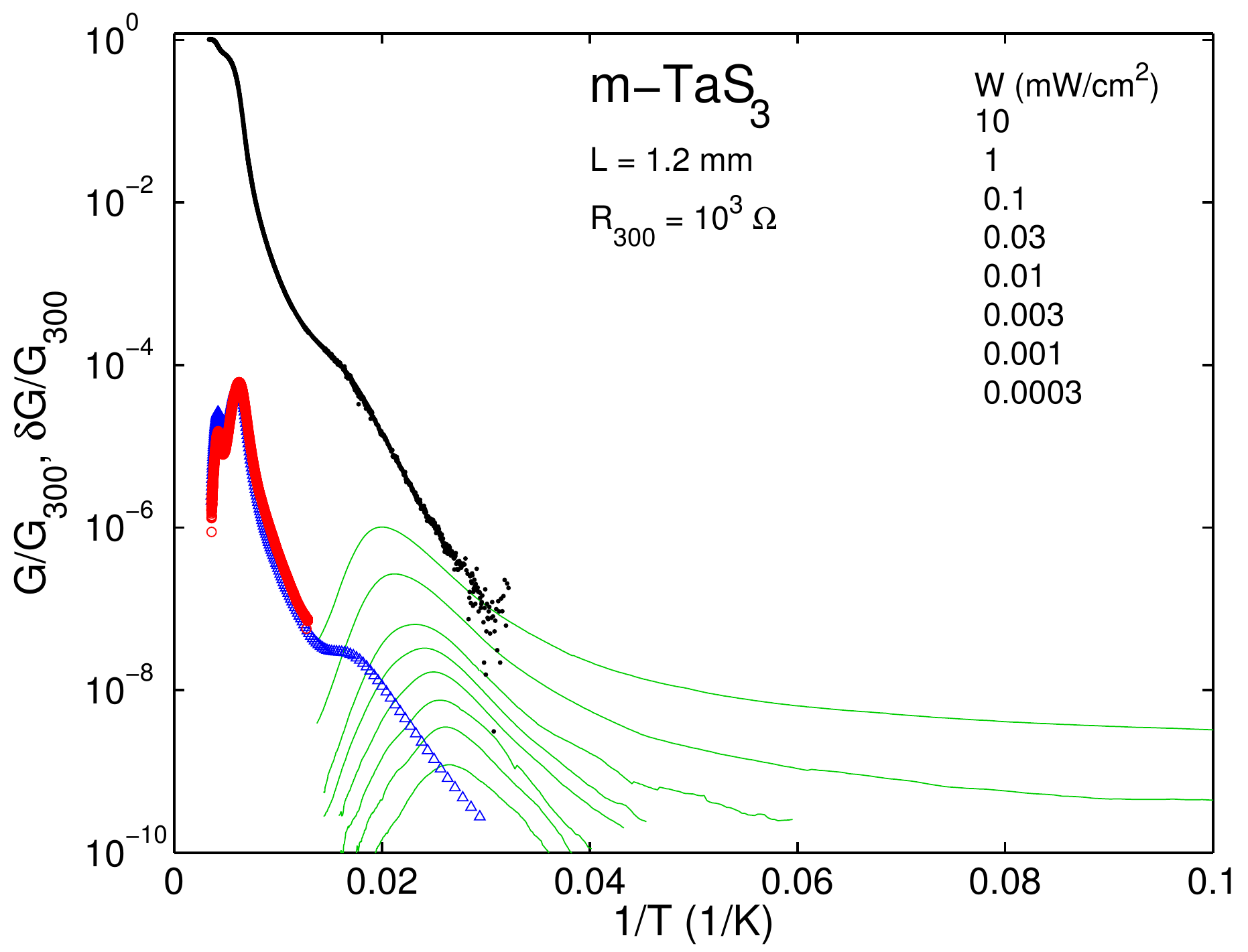}
\caption{Temperature dependences of: conductance of a thick sample of monoclinic TaS$_3$ (points);  $\delta T(dG/dT)$  (triangles); bolometric response at $W=10$~mW/cm$^2$ (circles); a set of photoconductance dependences obtained at $W$ values listed in the figure (lines). All the curves are normalized to $G_{300}$.}
\label{fig:monobolo}
\end{figure}

Fig.~\ref{fig:monobolo} demonstrates the difference between the bolometric and photoconduction responses in CDW conductors. It shows a typical set of temperature dependences of the photoconductance  together with the temperature dependence of the ohmic conductance, $G(T)$, and its derivative, $dG/dT$, of a thick {\it m-}TaS$_3$ sample. Bolometric response, being proportional to the illumination-induced temperature increase  ($\delta T=3.4$~mK in this particular case), follows $(dG/dT)\delta T$ dependence, whereas the photoconductance depends on the temperature in a very different way.  It is clear from this figure, that a good thermal contact is critical for photoconduction study. We also found, that the bolometric response is practically undetectable in much thinner samples. Note that $dG/dT$ drops dramatically in the low-temperature region, whereas photoconductance remains almost constant.
As a consequence, it is very difficult to study bolometric response of Peierls conductors at helium temperatures.
It seems therefore that the first photoconduction spectrum of q-1D conductor was actually obtained by Herr, Minton and Brill in 1986 in their low-temperature bolometric study \cite{bolofoto}.

%\section{Characteristic times, relaxation effects}
Photoconduction in the studied CDW conductors is detectable in the low-temperature range only at $T\lesssim T_P/2$. Temperature dependence of photoconduction measured at fixed illumination intensity depends mostly on the temperature dependence of the relaxation time of the nonequilibrium current carriers, $\tau$. In its turn, the relaxation time can be obtained from direct measurements of temporary response of photocurrent after switching the illumination on or off. The photoresponse kinetics was measured in blue bronzes by using the laser excitation \cite{ogawa} and in {\it o}-TaS$_3$ with LED excitation \cite{zzminajetpl}. In both cases the millisecond time scale of the characteristic time, $\tau$, was observed. It is known now that:
\begin{enumerate}

\item $\tau$ depends on the voltage applied and decreases with the voltage when it exceeds the threshold one for onset of nonlinear conduction \cite{ogawa}
\item $\tau$ follows the activation law at $T\gtrsim T_P/3$ \cite{zzminajetpl}
\item $\tau$ depends on the light intensity in a way which corresponds to the collisional recombination mechanism \cite{zzminajetpl}, intensity-dependent relaxation times are observed in the low-temperature region at $T\lesssim T_P/4$ \cite{zzminajetpl,zzmprl,zzminajphys}
\item long-time logarithmic-type relaxation  appears in the low-temperature region at $T\lesssim T_P/4$ \cite{zzminajphys}
\end{enumerate}
%The difference between the femtosecond spectroscopy measurements \cite{fsresponse} where picosecond-scale responses were observed comes from XXXX orders bigger level of excitation used there.

Relatively long relaxation time observed in photoconduction measurements allows to consider the current carriers (electrons and holes) as well-defined physical objects justifying thereby the usage of the semiconductor model \cite{APZZJETP} for description of many properties of the Peielrs conductors. Note, however, that the initial and final energy states of optically excited
electrons and holes may be different due to intrinsically high polarizability of crystal lattices of studied q-1D materials \cite{solitons_theor}.

%The most general conclusion obtained from the study is the ?????

\subsection{Effect of illumination on CDW kinetics}
There is very interesting effect of illumination on the CDW kinetics. This effect was observed in the blue bronze, K$_{0.3}$MoO$_3$, where illumination-induced growth of the threshold field was found \cite{ogawa}. The effect was then reproduced and studied in details in  {\it o}-TaS$_3$ \cite{zzminajetpl,zzminajphys}. It was shown that the effect results from the illumination-induced change of the screening which affects in its turn the elastic modulus of the CDW. The observed dependence, $E_T\propto G(W)^{1/3}$, corresponds to 1D pinning of the CDW.

\begin{figure}
\includegraphics[width=7.5cm]{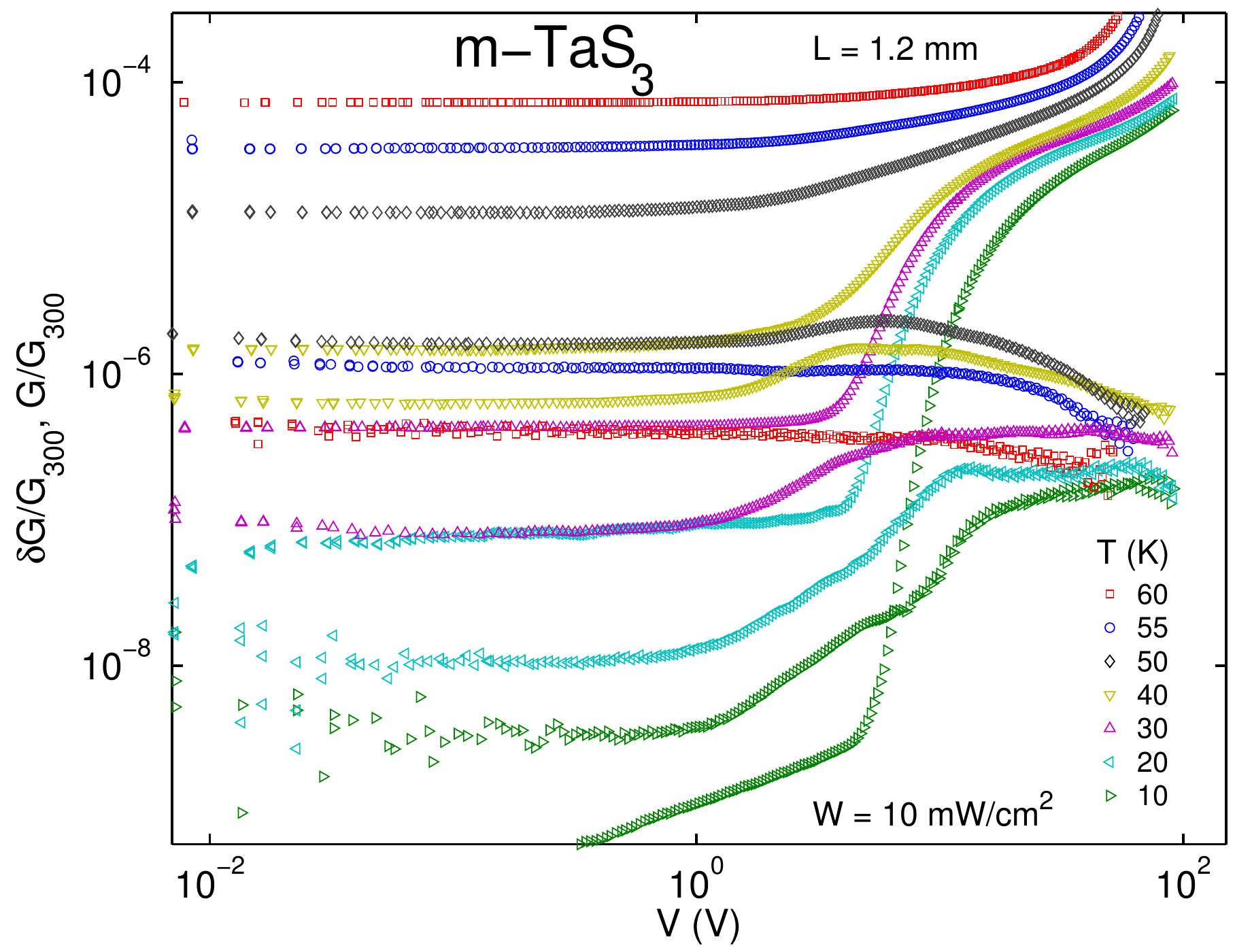}
\caption{Voltage dependence of the dark conductance (upper set of data) and photoconductance at $W=10$ mW/cm$^2$(lower set of data) of a {\it m}-TaS$_3$ sample at different temperatures.}
\label{fig:monovax}
\end{figure}

Slow CDW motion (CDW creep) at $E\lesssim E_T$ suppresses phococonduction \cite{zzminajetpl}. Negative photoconduction ($\delta G<0$) is observed at $E>E_T$ \cite{zzminajetpl,zzminajphys,ogawa} due to illumination-induced growth of $E_T$.

Surprisingly, a growth of photoconductance with the voltage applied can however be observed in some samples in the low-temperature region in the nonlinear conduction regime. In particular, such a growth was observed in both orthorhombic (see, e.g. Fig.~4 in Ref.~\cite{zzmnzreview}) and monoclinic modifications of TaS$_3$ (Fig.~\ref{fig:monovax}, see also \cite{minahere}).
Qualitatively, such a complex behavior corresponds to the competition between at least two physical mechanisms working in opposite directions. The first one is a mechanism of spatial separation of electrons and holes which complicates their recombination and is related to fluctuations of the chemical potential \cite{zzminajphys}. This mechanism smoothly disappears when the CDW starts to move in the electric field. The second one requires the presence of electronic states (impurities, defects, excitons) which are capable to capture electrons and provide their recombination, have relatively small excitation energy and can be depleted or destroyed by the electric field $10-100$~V/cm.
Electric-filed dependent energy states were really observed in the photoconduction spectra of {\it o}-TaS$_3$ \cite{nzzjetpl}, but their origin is still not entirely clear.

\section{Photoconduction spectroscopy}

\subsection{Fundamental edge and van Hove singularities}
\begin{figure}
\includegraphics[width=7.5cm]{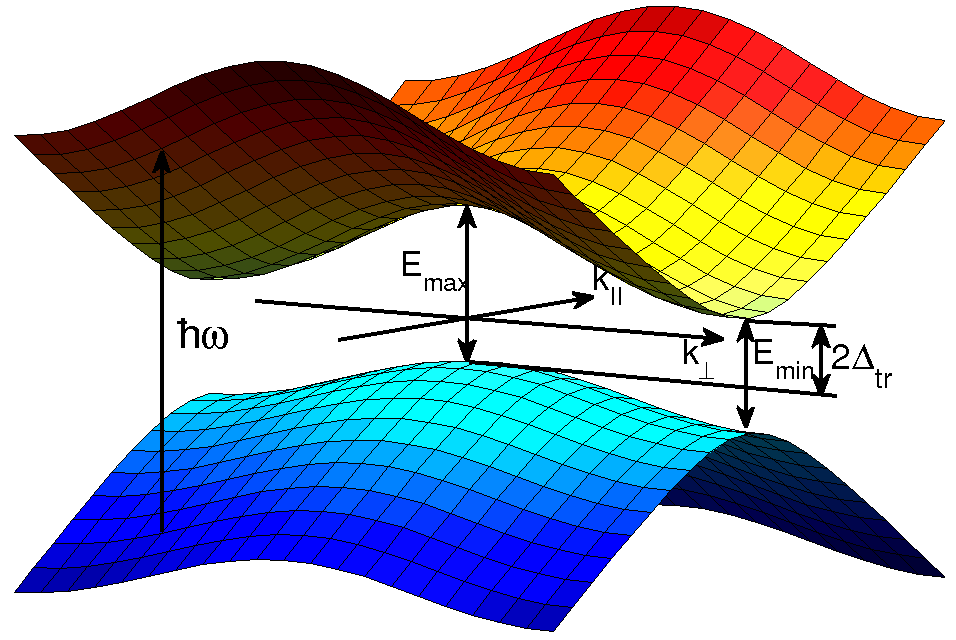}
\caption{3D view of the suggested modulation of the energy bands in the vicinity of the Peierls gap which separates the valence band $E_v({\bf k})$ (blue) and the conduction band $E_c({\bf k})$ (yellow-red colors). Direct optical transitions over the Peierls gap are vertical, $\hbar\omega = E_c({\bf k})-E_v({\bf k})$.}
\label{fig:3D}
\end{figure}

One of the most useful application of the photoconduction spectroscopy is determination of the optical energy gap value. In addition, a number of other features may be resolved by means of photoconduction spectroscopy. Imperfect nesting of the CDW leads to a periodic (in $k$-space) modulation of both the bottom, $E_v$, and the top, $E_c$, of the Peierls gap \cite{yamaji}. It is reasonable to suggest that the value of the gap, $E_c- E_v$, is also modulated in the directions perpendicular to the chains (see Fig.~\ref{fig:3D}). The physical origin of such a modulation is a noticeable curvature of the initial  $E(k_\parallel)$ dependence on the Peierls gap scale. Neglecting this curvature ($v_F={\rm const}$) results in $E_c- E_v$ to be independent of $k_\perp$  \cite{yamaji}.

\begin{figure}
\includegraphics[width=7.5cm]{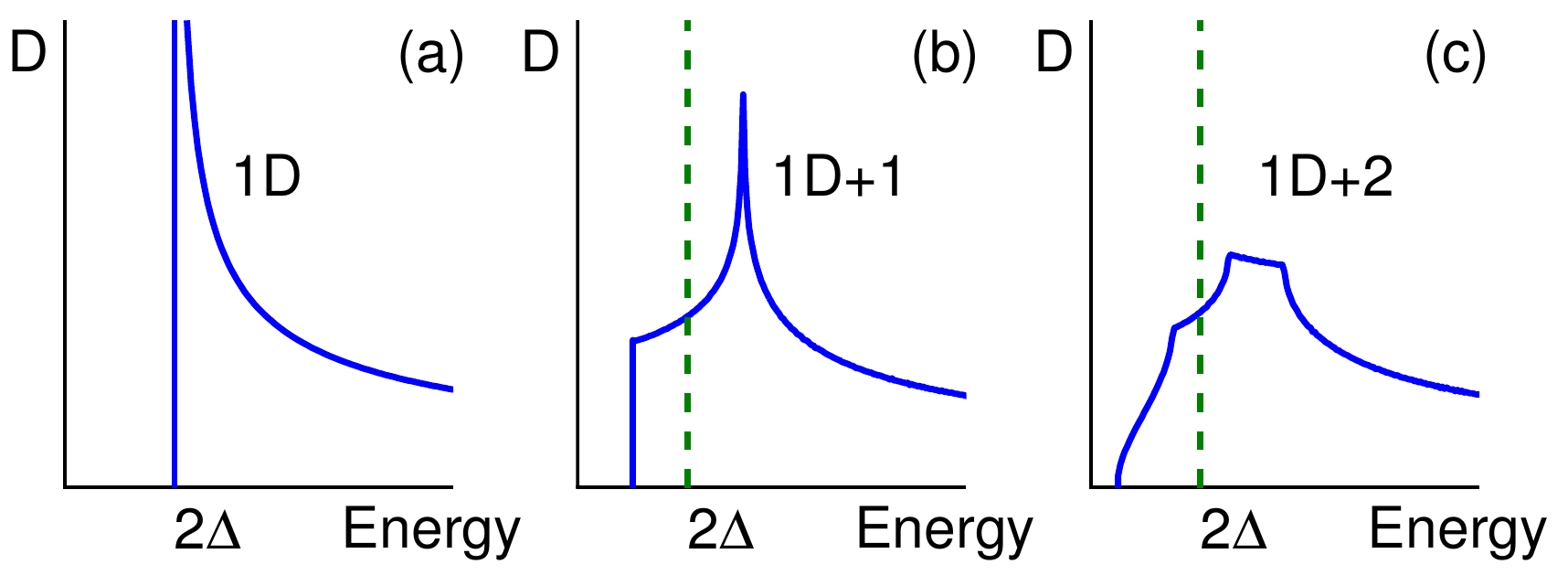}
\caption{Density of states for direct optical transitions over the Peierls gap in: (a) 1D case (no Peierls gap modulation), (b) 1D + 1 case with the Peierls gap modulated in one direction, (c) 1D + 2 case with the gap modulated in two directions.}
\label{fig:123D}
\end{figure}

For our purposes it is enough to consider the following cases: 
\begin{itemize}
\item[(a)] 1D case (no Peierls gap value modulation)
$$E_c(k)-E_v(k)=2\sqrt{\Delta^2+ (\hbar \delta k v_F)^2}\approx 2\Delta + (\hbar \delta k)^2/2m^*,$$ 
where $\delta k = |k| -k_F$, $k_F$ is the Fermi wave vector; 

\item[(b)] 1D + 1 case with the Peierls gap value modulated in one transverse direction, $$E_c({\bf k})-E_v({\bf k})=2\Delta + (\hbar \delta k_\parallel)^2/2m^*+\varepsilon_1\cos(\pi k_{\perp}/a);$$  

\item[(c)] 1D + 2 case with the Peierls gap value modulation in two transverse directions, 
\begin{equation}
\begin{split}
E_c({\bf k})-E_v({\bf k})=
2\Delta + (\hbar \delta k_\parallel)^2/2m^*+\\
+\varepsilon_1\cos(\pi k_{\perp 1}/a)+\varepsilon_2\cos(\pi k_{\perp 2}/b),
\end{split} 
\label{eq:model}
\end{equation}
where $a$ and $b$ are energy gap modulation periodicities in the directions normal to the chains, and $k_\parallel$, $k_{\perp 1}$ and $k_{\perp 2}$ are the components of the wave vectors along the chains and in the perpendicular directions. 
\end{itemize}
Then  $2\Delta_{opt}\equiv \min(E_c-E_v)= 2\Delta - |\varepsilon_1|-|\varepsilon_2|$ is the optical gap value of the present model, whereas $2\Delta_{tr}\equiv \min(E_c)-\max(E_v)$ is the transport gap. The respective optical densities of states are shown in Fig.~\ref{fig:123D}. Though the general expectation for the Peierls conductors is the bare 1D spectrum with the only van Hove singularity (Fig.~\ref{fig:123D}(a)), the presence of the modulation may make its shape to be very dissimilar owing to appearance of additional singularities (Fig.~\ref{fig:123D}(b),(c)). Respective photoconduction spectra may differ from shown in Fig.~\ref{fig:123D}. First of all, fluctuations of the order parameter which are intrinsic to q-1D systems partially smear out the van Hove singularities of the spectrum. We assume here Gauss fluctuations with a root mean square value $\delta\varepsilon$. In addition, the transport gap may be not direct (Fig.~\ref{fig:3D}). In this case a small tail appears inside the optical gap due to indirect phonon-assistant transitions. An additional contribution to the tail may also be a consequence of the Franck-Condone principle \cite{solitons_theor}. 

\begin{figure}
\includegraphics[width=7.5cm]{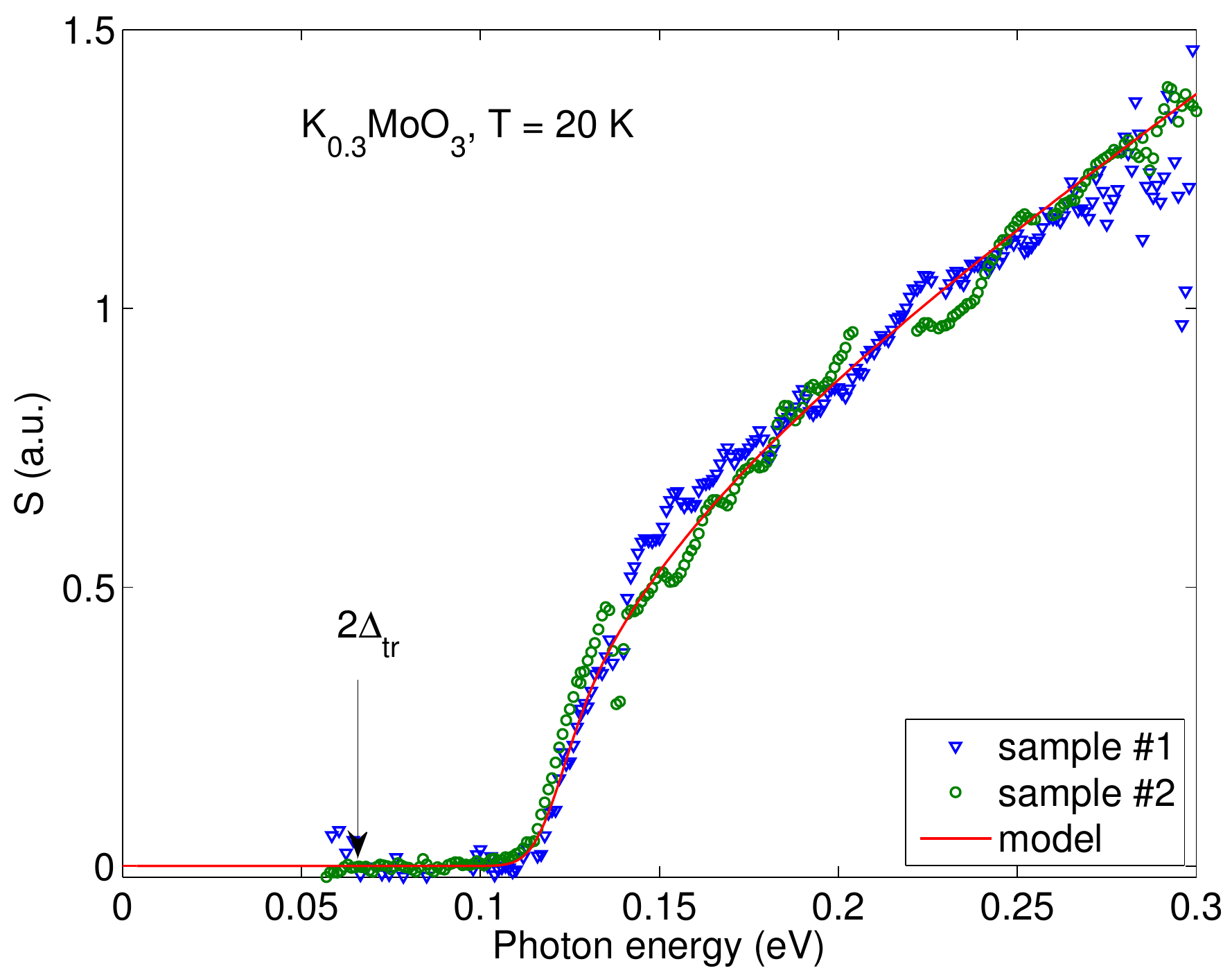}
\caption{Photoconduction spectra of two samples of K$_{0.3}$MoO$_3$ at $T=20$~K (data of Ref.~\protect\cite{zzmnzreview}) and their fit by Eq.~\ref{eq:model}. Fitting parameters: $\varepsilon_1= 0.25$~eV and $\varepsilon_2=0.35$~eV, $2\Delta_{opt} = 0.119$ eV, $\delta\varepsilon= 6$~meV.}
\label{fig:dosBB}
\end{figure}

Let us analyse now the photoconduction spectra data available up to now. On our experience,  photoconduction in the blue bronze K$_{0.3}$MoO$_3$ is well reproducible thought it is very complicated in measurements because of femtoampere-scale photoresponse in the narrow optimal temperature range around 20 K.  Fig.~\ref{fig:dosBB} shows the best fit of the experimental results (Ref.~\cite{zzmnzreview}) with the 1D+2 DOS similar to one shown in Fig.~\ref{fig:123D}(c). We can see that this simple model fits the data fairly well. Note that the parameters obtained imply very strong modulation of the gap value in $k$-space and, therefore, in the real space. The presence of such modulations can be verified experimentally in ARPES and scanning-tunneling spectroscopy measurements respectively. 

\begin{figure}
\includegraphics[width=7.0cm]{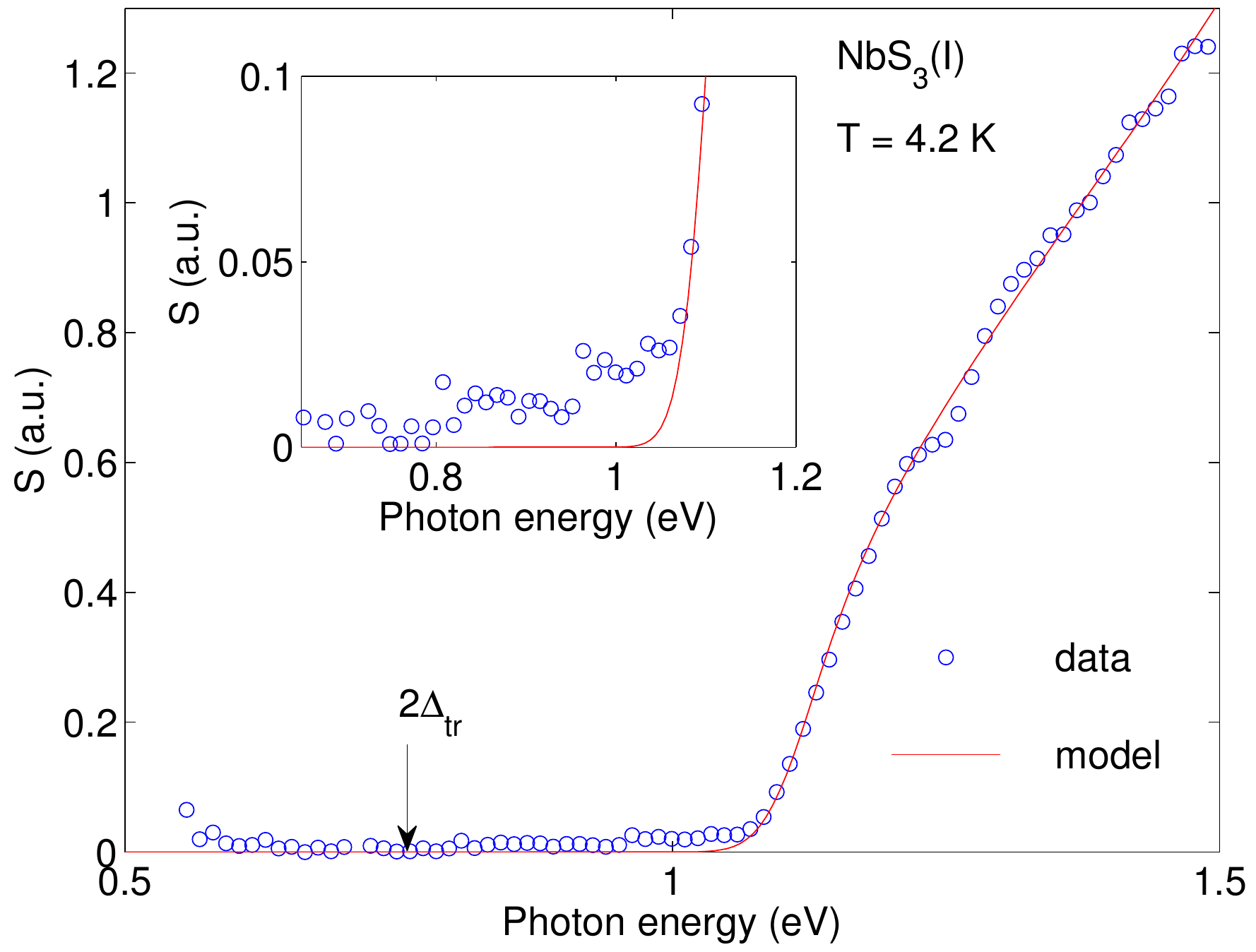}
\caption{Typical photoconduction spectrum of NbS$_3$ at $T=4.2$~K (circles, data of Ref.~\protect\cite{nbs3}) and its fit by Eq.~\protect\ref{eq:model}.  Fitting parameters: $\varepsilon_1= 0.6$~eV and $\varepsilon_2=0.3$~eV, $2\Delta_{opt} = 1.11$~eV, $\delta\varepsilon=30$~meV. Inset shows the tail of states near the band edge.}
\label{fig:dosnbs3}
\end{figure}

Fig.~\ref{fig:dosnbs3} shows the typical low-temperature photoconduction spectrum of NbS$_3$(I) sample. Again, the fitting works well for this compound. 

\begin{figure}
\includegraphics[width=7cm]{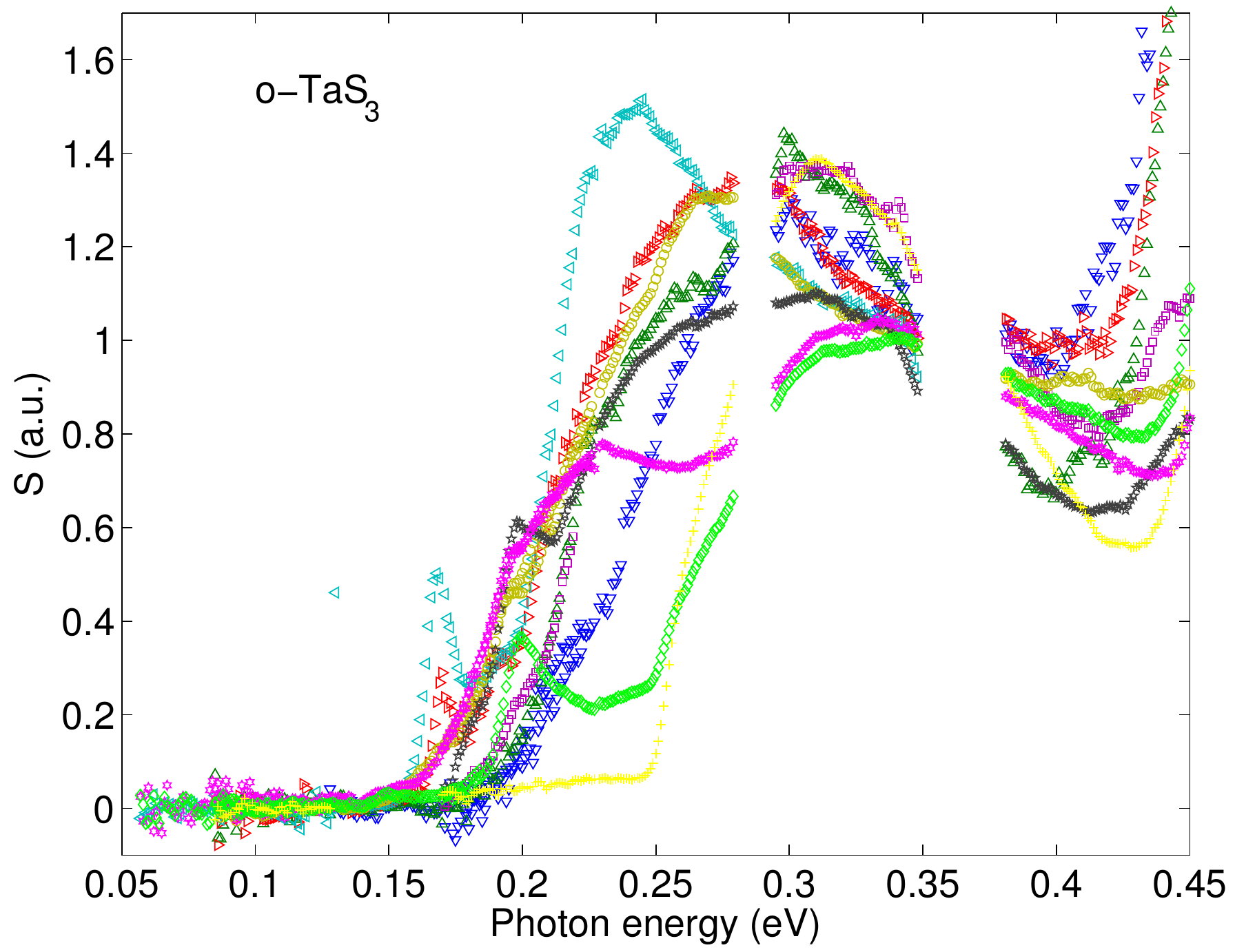}
\caption{A set of photoconduction spectra of {\it o}-TaS$_3$ samples of different purity.}
\label{fig:tas3all}
\end{figure}

In {\it o}-TaS$_3$, an observable Peierls gap edge varies noticeably from sample to sample. Even crystals from the same batch or different fragments of the same crystal may have substantially different energy spectra. Fig.~\ref{fig:tas3all} shows a set of spectra of samples from different batches. In contrast with highly reproducible NbS$_3$(I) and K$_{0.3}$MoO$_3$ data, the spectra parameters have very wide distribution. 

\begin{figure}
\includegraphics[width=7cm]{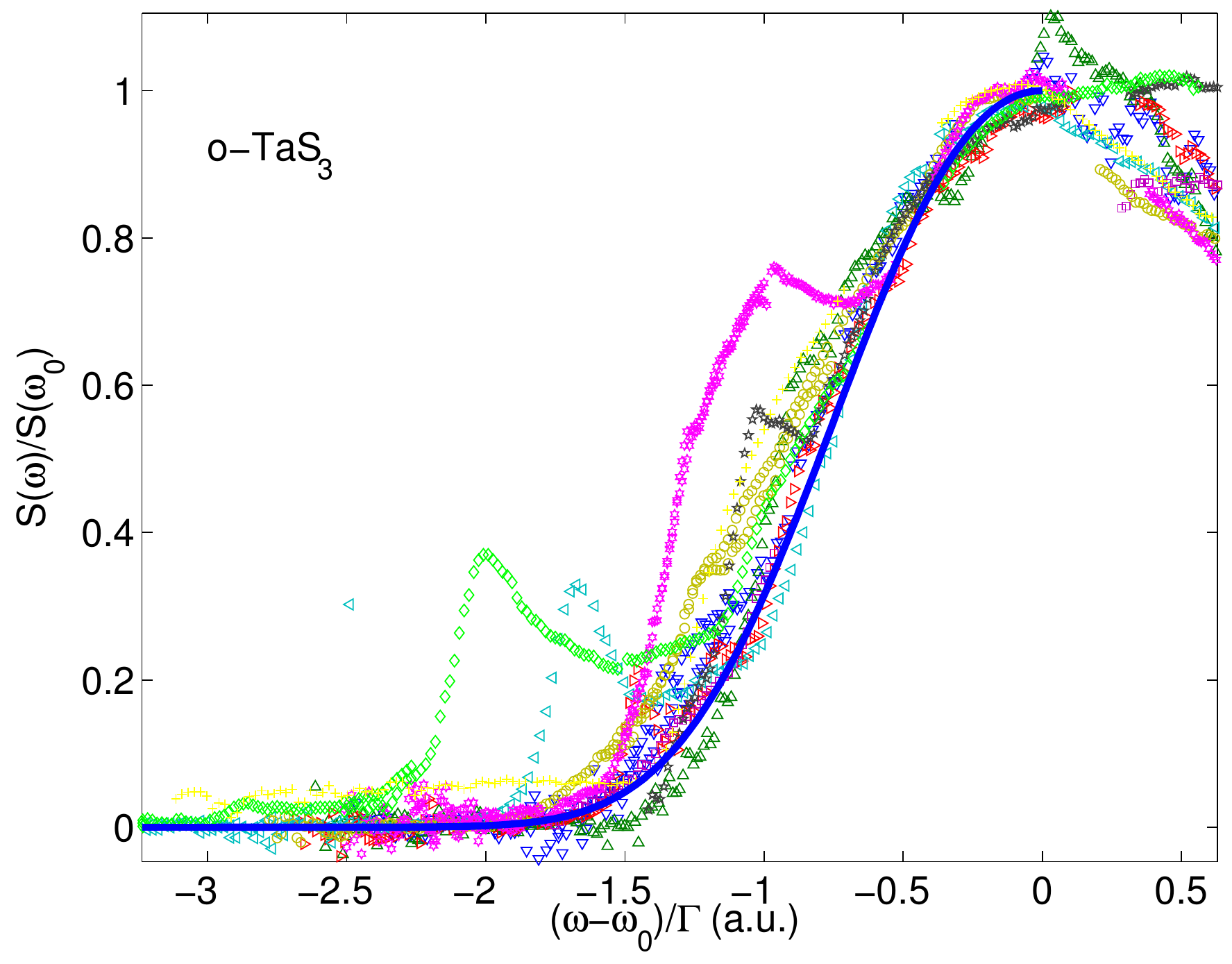}
\caption{Scaling plot of experimental data shown in Fig.~\protect\ref{fig:tas3all}. For each curve the  parameter $\Gamma$ was choosen to give the best fit to the theoretical dependence described by equation \protect\ref{eq:mckenzy} (solid line).}
\label{fig:tas3allfitted}
\end{figure}

The effect of impurities and fluctuations on the Peierls gap edge was studied theoretically in Ref.~\cite{kim}. It was shown that the low-energy tail ($\omega\leq 2\omega_0$) of the optical conductance follows the universal law 
\begin{equation}
\sigma = \sigma_0\exp{\left[-c_1\left|\frac{\omega_0-\omega}{\Gamma}\right|^2-c_2\left|\frac{\omega_0-\omega}{\Gamma}\right|^3\right]},
\label{eq:mckenzy}
\end{equation}
where $c_1$, $c_2$ are numerical coefficients, $\omega_0$ is the peak position, $\Gamma$ is a fluctuation scale. That means that $\sigma/\sigma_0$ is a universal function of $(\omega_0-\omega)/\Gamma$. This dependence was successfully applied to description of the experimental data of KCP(Br) and {\it trans}-(CN)$_x$ (see \cite{kim} and references therein). Fig.~\ref{fig:tas3allfitted} shows the same photoconduction data as in Fig.~\ref{fig:tas3all} rescaled in accordance with Eq.~\ref{eq:mckenzy}. The scaling works reasonably well for {\it o}-TaS$_3$, especially for impure samples or samples subjected to a large number of thermal circling or long exposition at the normal conditions (see also \cite{venerahere}). It means that the wide distribution of the gap edge observed at least partially is consequence of the impurity-induced order parameter fluctuations.

\begin{figure}
\includegraphics[width=7cm]{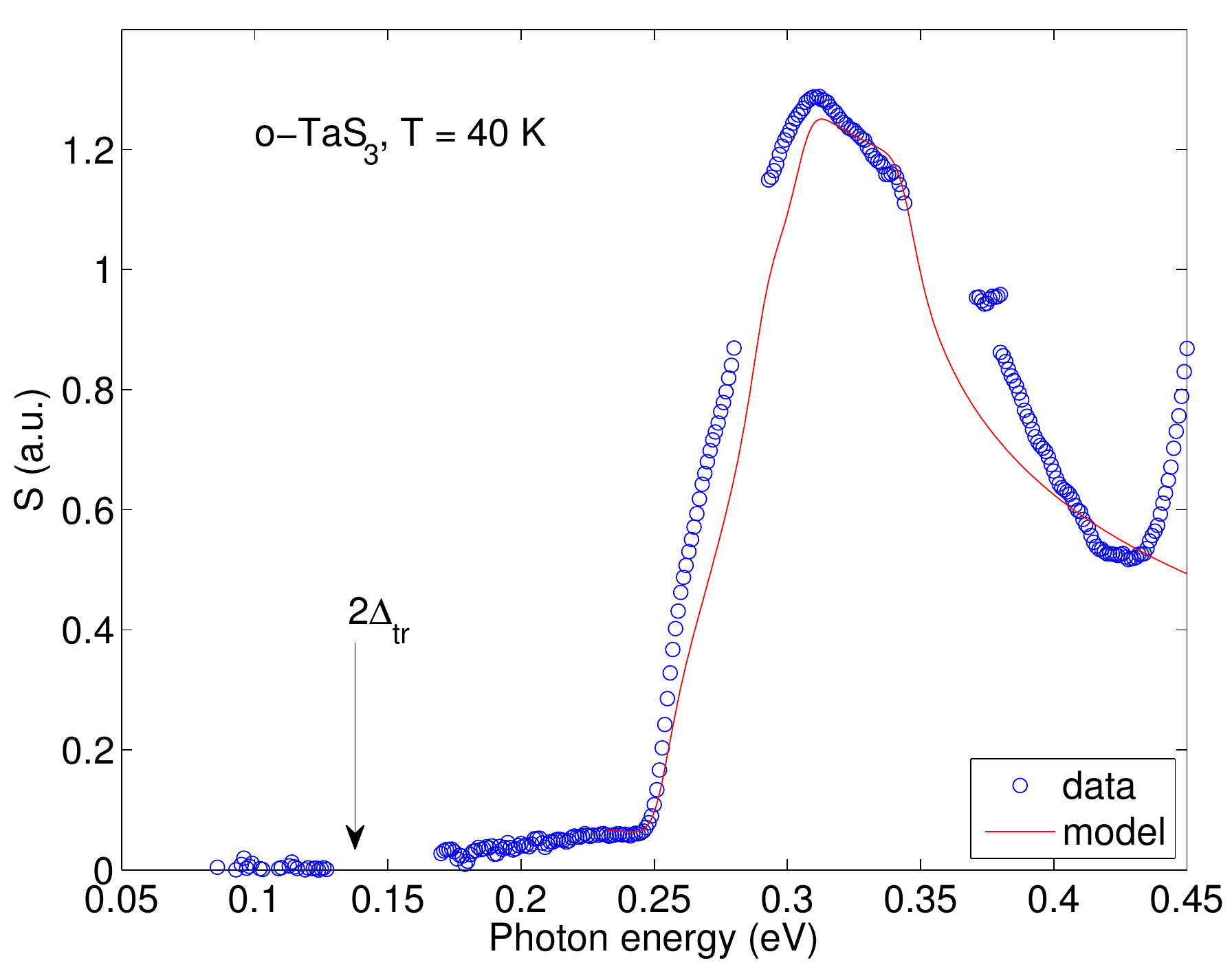}
\caption{Photoconduction spectrum of a freshly grown and prepared {\it o}-TaS$_3$ sample (data of Ref.~\cite{venerahere}) and its fit by Eq.~\ref{eq:model} with the parabolic dispersion law. Fitting parameters:  $2\Delta=0.253$~eV, $\varepsilon_1=27$~meV and $\varepsilon_2=19$~meV, $\delta\varepsilon=4$~meV.}
\label{fig:tas3fresh}
\end{figure}

Fig.~\ref{fig:tas3fresh} shows a photoconductance spectrum of a freshly synthesized and prepared  {\it o}-TaS$_3$ sample. A sharp Peierls gap edge singularity (at 0.25~eV) and a cusp at 0.3 eV  cannot be described in the framework of the fluctuation theory \cite{kim}. Instead, a simple model of the Peierls gap modulation (Fig.~\ref{fig:123D}) works much better for this sample. It is even able to catch all the observed van Hove singularities. Some deviation from the model is seen at energies above 0.35 eV and corresponds to non-sine modulation of the energy spectra in the vicinity of the next band. ``Ageing'' of this sample at $T=120^o$C transforms its spectrum into the ordinary one \cite{venerahere}.

\begin{figure}
\includegraphics[width=7cm]{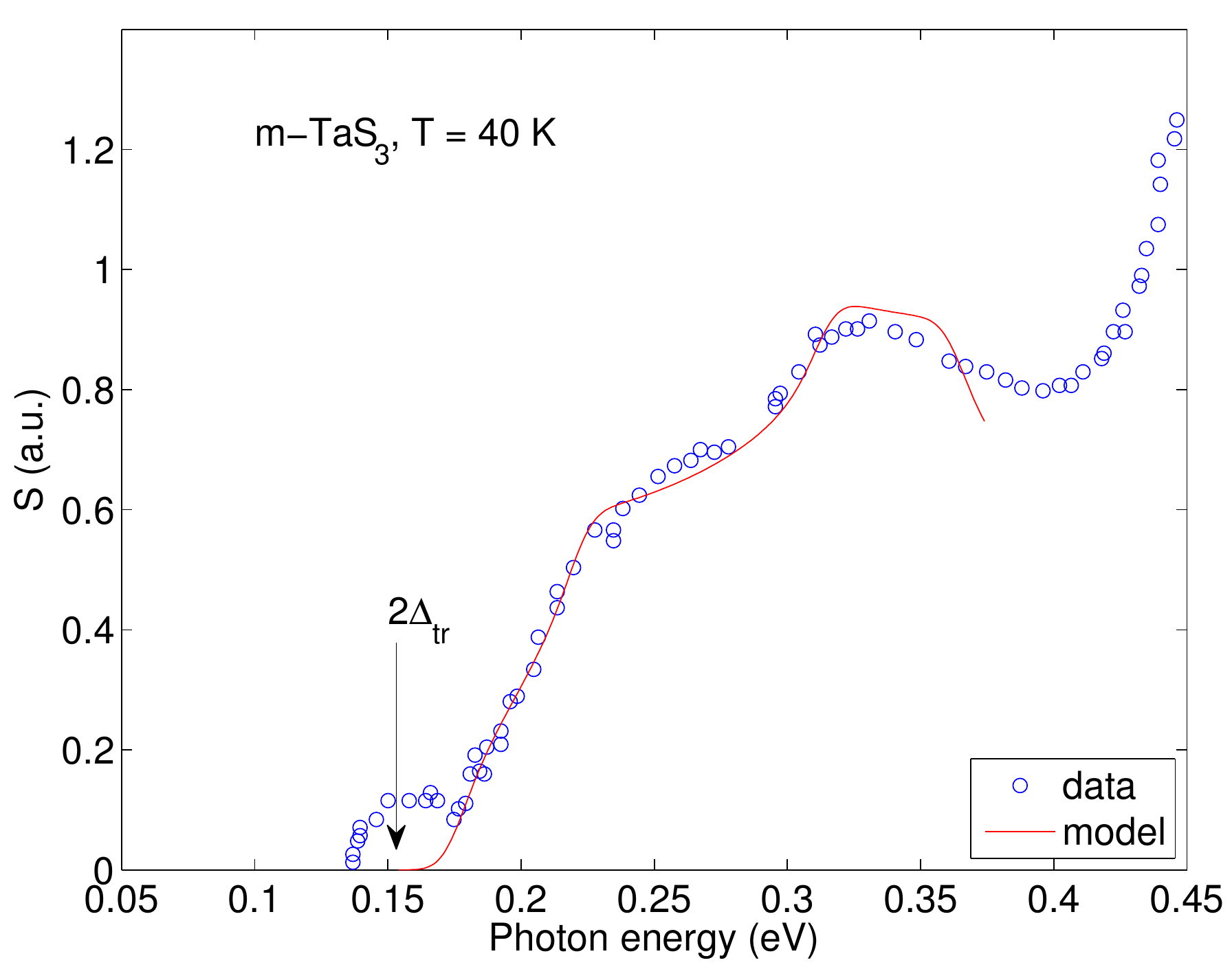}
\caption{Photoconduction spectrum of {\it m}-TaS$_3$ sample at $T=40$~K (data of Ref.~\cite{minahere}) and its fit by Eq.~\ref{eq:model} with the parabolic dispersion law. Fitting parameters: $2\Delta=0.18$~eV, $\varepsilon_1=68$~meV and $\varepsilon_2=24$~meV, $\delta\varepsilon=6$~meV.}
\label{fig:mono}
\end{figure}

Photoconduction spectrum of {\it m}-TaS$_3$ sample is shown in Fig.~\ref{fig:mono}. Two low-energy features can be distinguished, the edge at 0.14 eV and the dip at 0.18 eV. As  0.14 eV is substantially smaller than the transport gap, so the energy region 0.14 eV - 0.18 eV corresponds to the states which do not participate in the charge transport, and the gap edge corresponds actually to the dip. The model of the gap modulation is able to fit the low-energy part of the curve, but deviate from the experimental data at higher energies $\hbar\omega \gtrsim 0.4$~eV, similar with {\it o}-TaS$_3$ data (see Fig.~\ref{fig:tas3fresh}). Polarization study shows that the peak at $\hbar \omega\approx 0.15$~eV appears only for the light polarised along the chains, whereas the band structure is visible in both polarizations with larger response for the perpendicular one.

In the framework of the semiconductor model \cite{APZZJETP}, the chemical potential level sits near the middle of the Peielrs gap and the possible deviation does not exceed a few $kT$. Therefore the transport gap value can be estimated as the doubled conductance activation energy. Table~\ref{tab:gaps} shows a summary of the obtained results. One can see that the optical gap of all the studied materials exceeds the transport one.  The difference between the optical and transport gaps may have two sources. The first one is a consequence of the Franck-Condon principle: the initial and final energy states of optically excited electrons and holes may be different due to high polarizability of crystalline lattices of studied q-1D materials \cite{solitons_theor}. The second one is indirect band structure in the studied materials. In both cases a low-energy tail inside a direct optical gap is expected. Indeed, such tails are seen in all studied materials except for {\it m}-TaS$_3$ where it is masked by the wide peak at 0.15 eV.

We would like to note surprisingly small level of fluctuations required for the best fitting of the spectra. The typical rms value $\delta\varepsilon \sim 5$~meV amounts only a few percent of the respective optical energy gap value and is comparable with the spectral resolution of our setup. Such fluctuations are responsible  only for insignificant smoothing of the energy gap features and leave the van Hove singularities of all the studied compounds to be observable.

\begin{table}
\caption{Comparison of the transport and optical data.}
\begin{tabular}{|c|c|c|c|}
\hline 
\rule[-1ex]{0pt}{2.5ex} material & activation &transport & optical\\ 
\rule[-1ex]{0pt}{2.5ex}   & energy & gap & gap \\ 
\hline 
\rule[-1ex]{0pt}{2.5ex} o-TaS$_3$ & 800 K&0.137 eV & 0.25 eV \\ 
\hline 
\rule[-1ex]{0pt}{2.5ex} m-TaS$_3$ & 900 K~\cite{comment}& 0.155 eV &0.18 eV\\ 
\hline 
\rule[-1ex]{0pt}{2.5ex} K$_{0.3}$MoO$_3$ & 326 K & 0.056 eV &0.119 eV \\ 
\hline 
\rule[-1ex]{0pt}{2.5ex} NbS$_3$(I) & 4300 K &0.75 eV & 1.11 eV \\ 
\hline 
\end{tabular} 
\label{tab:gaps}

\end{table}

\subsection{Impurities, defects and various excitations}
In addition to the optical gap measurements the photoconduction may be used for identification of the energy levels inside the Peierls gap. Peaks in photoconduction spectra may result from impurities, excitons, solitons, polarons, macroscopic defects of the crystalline lattice (dislocations, grain boundaries, packing faults {\it etc.}) and similar defects of the CDW. Such states may appear even at energies above the optical gap owing to complexity of the band structure of the Peierls conductors.

The first photoconduction spectra measurements gave evidences of existence of relatively narrow peaks in {\it o}-TaS$_3$ \cite{nzzjetpl}, the amplitude of some of them being dependent on a relatively small electric field $E\sim 10-100$ V/cm \cite{nzzjetpl,nbs3}.. Somewhat similar peaks were also found in NbS$_3$(I) in the middle of the optical gap at 0.6 eV and near the temperature-dependent edge at 0.9~eV \cite{nbs3} and quite recently in m-TaS$_3$ \cite{minahere}. Similar peaks are also observed in the bolometric spectra of o-TaS$_3$ \cite{bolobrill,boloitkis}. 

The spectral peaks observed in {\it o-}TaS$_3$ near the optical gap edge (0.25 eV \cite{venerahere}) have asymmetrical shape (Fig.~\ref{fig:tas3allfitted}) and relatively large amplitudes, comparable with those of the main maximums. Numerous attempts of identification of such peaks with impurities in crystals doped by Ti, Nb and In gave no clear results \cite{zzmnzreview}. Instead, it was found that low-purity samples have less peaks, whereas the peaks are always present in very clean one. It is very unlikely, therefore, that the peaks correspond to impurity states, their relations to excitonic or polaronic states look much more plausible. For instance, excitons in the CDW conductors may have bigger oscillator strength than the interband transitions \cite{abe}, so their contribution may be even dominating in very pure crystals. Further study is required for  identification of the peaks. 

\begin{figure}
\begin{center}
\includegraphics[width=6.7cm]{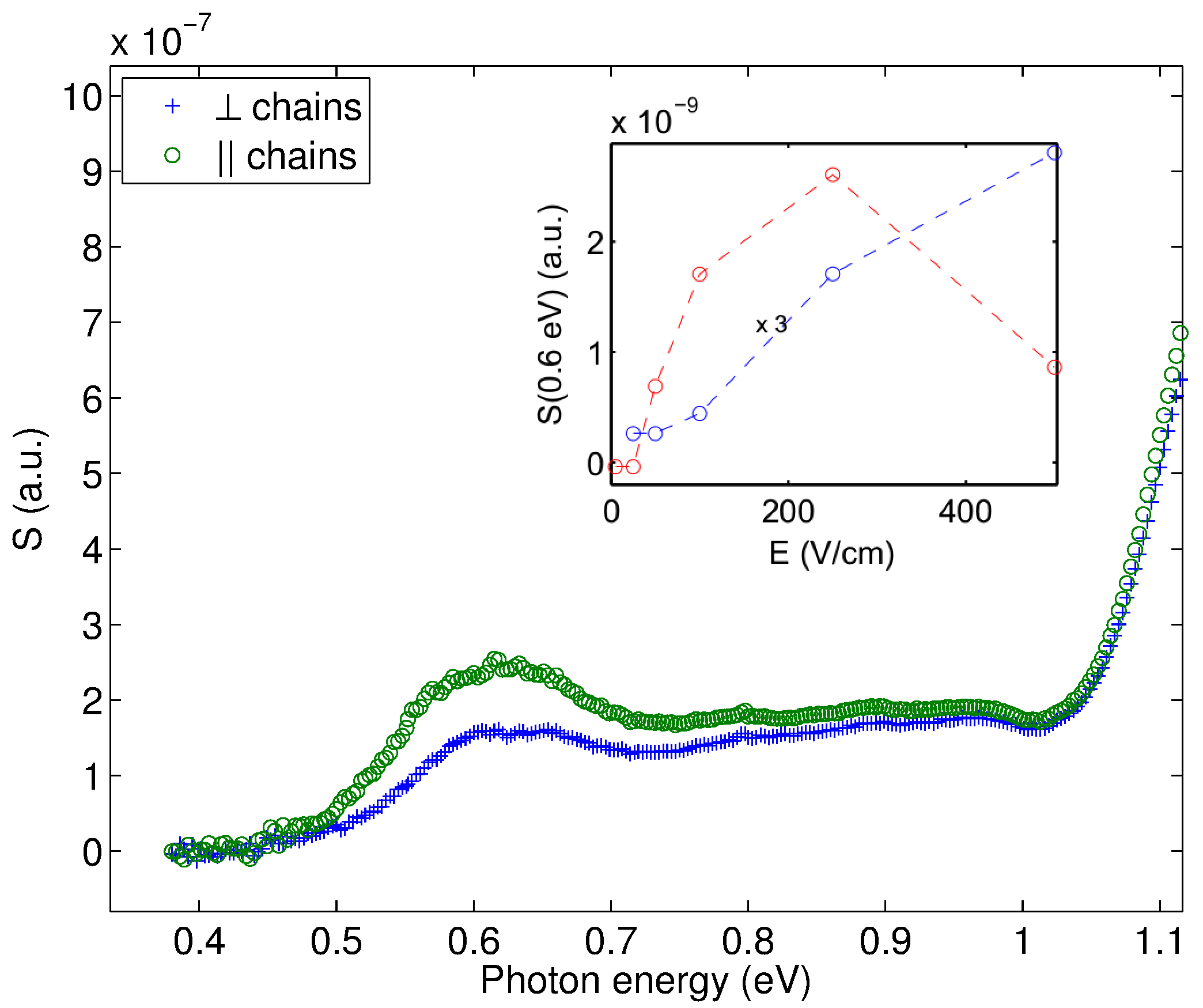}
\end{center}
\caption{Polarization-dependent photoconduction spectra of NbS$_3$(I) in the region of the electric-field dependent mid-gap peak. $E=330$ V/cm, $T=78$ K, polarizer wavelength range 2-30~$\mu$m. Inset: Dependence of the amplitudes of the mid-gap peaks on the electric field in two samples of NbS$_3$(I), data of Ref.~\protect\cite{nbs3}.}
\label{fig:nbs306peak}
\end{figure}

Existence of solitons and their possible contribution into physical properties of the Peielrs conductors is one of the most long-standing problem.  Amplitude solitons are expected to produce electron states near the middle of the Peierls gap \cite{solitons_theor}. As the concentration of such states is equal to the concentration of solitons, so their excitation by additional illumination or current injection results to increase of the soliton spectral peak. Surprisingly, no such peaks were observed in  photoconduction spectra of the ``classical'' CDW conductors such as K$_{0.3}$MoO$_3$ and both phases of  TaS$_3$, whereas the mid-gap peak at 0.6~eV demonstrates exactly this type of behavior in all studied samples of NbS$_3$(I) \cite{nbs3} (see Fig.~\ref{fig:nbs306peak}). Polarization study shows that the peak is observed for the light polarised along the chains.  
% That means parallel to the chains orientation of the respective dipole moment.
Note that this material can be considered as a crystalline analogue of polyacetylene where similar mid-gap states are observed and ascribed to soliton states \cite{solitons_exper}.

\section{Conclusion}
The results described above demonstrate that photoconduction study as a method of experimental investigation of the charge-density wave conductors is fruitful and provides the new data, which are hardly available from other methods. In particular, photoconduction measurements  allow to observe the effect of carrier concentration on collective transport, to determine the relaxation time for nonequilibrium current carriers and to obtain various details of the energy structure.   

%\section{Acknowledgements}
{\bf Acknowledgements.} We are grateful to S.A. Brazovskii for useful discussions. Financial supports from RFBR and Department of Physical Sciences of RAS are acknowledged.

\end{document}